# Saturated Sodium Chloride Solution under an External Static Electric Field:  a Molecular Dynamics Study

Gan Ren[a), b)]  and Yanting Wang[a)]

[a)] *State Key Laboratory of Theoretical Physics, Institute of Theoretical Physics, Chinese Academy of Sciences, 55 East Zhongguancun Road, P. O. Box 2735 Beijing, 100190 China*

[b)] *University of Chinese Academy of Sciences, Beijing, 100190 China*

**ABSTRACT**

The behavior of saturated aqueous sodium chloride solutions under a constant external electric field ($E$) was studied by molecular dynamics (MD) simulation. Our dynamic MD simulations have indicated that the irreversible nucleation process towards crystallization is accelerated by a moderate $E$, but retarded or even prohibited under a stronger $E$, which can be understood by the competition between self-diffusion and drift motion. The former increases with $E$ resulting in the acceleration of the nucleation process, and the latter tears oppositely charged ions more apart under a stronger $E$ leading to the deceleration of nucleation. Moreover, our steady-state MD simulations have indicated that a first-order phase transition happens in saturated solutions only when the applied $E$ is below a certain threshold $E_c$, and the ratio of crystallized ions does not change with the electric field. The magnitude of $E_c$ increases with concentration, because larger clusters are easy to form in a more concentrated solution and require a stronger $E$ to dissociate them.

1. **Introduction**

Ion nucleation and crystallization in ionic solutions are important for various scientific areas, such as physical science,[1, 2] materials science,[3] and biology.[4] The ion nucleation process can be qualitatively understood by the classical nucleation theory (CNT),[5, 6] in which particles form clusters through a nucleation process, and crystallization takes place when a cluster grows large enough to exceed the size of a critical nucleus. In the CNT theory, the cluster size is regarded as the only order parameter to describe the nucleation process, and the change of the Gibbs free energy with cluster size $n$ can be expressed as

$$\Delta G(n) = -\alpha n + \gamma n^{2/3} \quad (1)$$

where $\alpha$ is the free energy difference between the metastable liquid phase and the stable solid phase, and $\gamma$ is the surface free energy of the interface between solid and liquid phases. The critical nucleus size and nucleation rate are two main quantities in the CNT to describe a nucleation process.

In ionic solutions, the instantaneous cluster formation due to the electrostatic interaction among ions is the precursor of an irreversible nucleation process. [7] The cluster formation in ionic solutions had been proposed a long time ago on the basis of analyzing the difference between theoretical prediction and experimental results,[8] but only recently did some experiments[9, 10] with enhanced experimental techniques observe instantaneous clusters formed in ionic solutions. On the other hand, the microscopic properties of ionic clusters and the details of a nucleation process in a solution can be better studied by computer simulation. With the aid of MD simulations, the cluster formation in NaCl solutions at ambient[11, 12] and supercritical[13] conditions were observed; Hassan[14, 15] explored the structural and dynamic properties of ion clusters in NaCl solutions; Zahn[16, 17] studied the stable structures of nuclei in different ionic solutions; and the nucleation at the interface of ionic solutions was also studied.[18] Different nucleation pathways in $CaCO_3$ solutions[19-21] were discovered by MD simulation, which has demonstrated that the nuclei are liquid-like chains formed by ions rather than the spherical shape assumed in CNT, and the cluster stability deviates from the CNT prediction. Furthermore, our MD simulations[7] have demonstrated that ion distributions in an ionic solution is homogeneous only from an ensemble-averaged point of view, and the instantaneously formed ion clusters due to thermal fluctuations are the precursor of an irreversible nucleation process leading to ion crystallization.

In the experiment, by applying a static external electric field $E$ to the ionic solution, Wien[22] observed the Wien effect that the conductance of weak electrolyte solutions increases with the intensity of $E$. Onsager and Kim[23] have explained theoretically the Wien effect from a Brownian motion viewpoint by showing that the dissociation constant for oppositely charged ions under $E$ is different from that without $E$ by a factor $\beta=1+kE+(\frac{1}{3})k^2E^2+\cdots$, where $k$ is a positive constant. Our previous calculations have also shown that ion clusters tend to be smaller in concentrated solutions when $E$ is applied.[24]

Since an applied $E$ destroys the homogeneity of an ionic solution, [24] an interesting question arises: how does the nucleation process in an ionic solution under $E$ differ from the homogeneous nucleation process without $E$? In addition, since a crystallized cluster in a saturated solution is more stable than a disordered cluster in a concentrated solution, another interesting question is: can a crystallized cluster be destroyed by a small $E$ and re-dissolve into the solution? Consequently, what will happen when $E$ is applied to an ionic solution which has already crystallized? Does there exists a phase transition point different from the case without an external $E$?

To answer the above questions, in this work, we investigate saturated sodium chloride (NaCl) solutions by MD simulation in two cases: a set of dynamic MD simulations (evolving dynamically towards the non-equilibrium steady-state) to study the influence of $E$ on the nucleation process, and a set of steady-state MD simulations to study the effect of $E$ to the thermodynamic properties of crystallized ionic solutions. Recently, MD simulations[25, 26] have shown that the nucleation occurs along many different pathways, so a single order parameter of cluster size is not enough to describe the nucleation process satisfactorily. Moreover, the critical nucleus size can be very different calculated by different methods.[27] Therefore, in this work, we mainly discuss the influence of *E* on the nucleation rate rather than the cluster or nucleus size.

The paper is organized as follows: section 2 contains a brief description of the simulation details and analysis methods, and simulation results are described in section 3, followed by section 4 including our conclusions and discussion.

## 2. Simulation details and analysis methods

In this section, we describe the details of our MD simulations, the heterogeneity order parameter used to quantify the nucleation process, and the bond order parameters to characterize different structures.

### 2.1. Simulation details

The CHARMM force field[28] for NaCl and the TIP3P[29] water model were used to model all the simulated ionic solutions. All ions and atoms are modeled as charged Lennard-Jones particles and the interactions between atoms and/or ions are described as

$$u(r_i, r_j) = 4\varepsilon_{ij}\left[\left(\frac{\sigma_{ij}}{r_{ij}}\right)^{12} - \left(\frac{\sigma_{ij}}{r_{ij}}\right)^{6}\right] + \frac{q_i q_j}{r_{ij}} \quad (2)$$

where $r_i$ is the *i*th atom (ion) position, $\sigma$ and $\varepsilon$ are the Lennard-Jones distance and energy constant, respectively, and $q_i$ is the partial charge of the *i*th atom (ion). The corresponding force field data are listed in Table 1. All our MD simulations were carried out with the GROMACS package[30, 31] and the system temperature was kept a constant by the Nosé-Hoover thermostat.[32, 33] The simulated systems contain 3000 water molecules and different pairs of Na$^+$ and Cl$^-$ corresponding to different concentrations above the saturation point, which was determined to be about 5 M by our previous MD simulations.[7] The periodic boundary conditions were applied to all three dimensions and the particle mesh Ewald algorithm[34] was employed to calculate the long-range electrostatic interactions.

Ions and H₂O molecules were put in a cubic box whose side length was first determined by a 5-ns constant *NPT* simulation at temperature $T = 300$ K and pressure $P = 1$ atm to determine the system density. A random configuration was chosen from a 2-ns *NVT* MD simulation at $T = 2000$ K as the initial configuration for a 250-ns dynamic *NVT* MD simulation procedure under $E$ at $T = 300$ K to investigate the nucleation process. Another 10-ns steady-state MD simulation under the same condition was also sampled to study the influence of $E$ on crystallized solutions which had been first run for 250 ns to allow the system to reach its non-equilibrium steady state before sampling. The time step for all MD simulations was 1 fs and the configurations were sampled every 1000 steps for data analysis.

**Table 1.** CHARMM force field parameters for Na⁺ and Cl⁻ and TIP3P water model parameters.

| Atom/ion | σ(nm) | ε(kJ/mol) | Charge(e) |
| --- | --- | --- | --- |
| Na⁺ | 0.243 | 0.196 | +1 |
| Cl⁻ | 0.405 | 0.628 | -1 |
| O | 0.315 | 0.636 | -0.834 |
| H | — | — | 0.417 |

### 2.3. Heterogeneity order parameter

The heterogeneity order parameter (HOP) was adopted to characterize the homogeneity of the simulated solutions. The HOP of an ion is defined as

$$\tilde{h}_i = \sum_{j=1}^{N} \exp\left(-r_{ij}^2 / 2\lambda^2\right) \quad (3)$$

where $r_{ij}$ is the distance between ions $i$ and $j$ corrected with the periodic boundary condition and $\lambda = L/N^{1/3}$, where $L$ is the side length of the cubic simulation box and $N$ is the total number of ions in this configuration. The HOP for an instantaneous configuration is defined to be the average HOP over all ions as

$$\tilde{h} = \frac{1}{N}\sum_{i=1}^{N}\tilde{h}_i \tag{4}$$

The HOP was initially defined to characterize the tail aggregation phenomenon in ionic liquids.[35] Similar to the radial distribution function (RDF), it can be used to characterize the structure of a system. A rigorous connection between the HOP and the RDF was given in ref [36]. In order to have a value close to zero when the ion distribution is nearly uniform, the reduced HOP is defined as

$$h = \tilde{h} - \tilde{h}_0 \tag{5}$$

where $\tilde{h}_0$ is the HOP for an ideally uniformly distributed system. The values of $\tilde{h}_0$ for different sizes are listed in ref 34.

### 2.3. Bond order parameter

The bond order parameters (BOPs)[37, 38] were employed to distinguish crystalline structures from the liquid structure. A bond is defined as the vector joining a pair of neighboring particles whose interatomic distance is less than a cutoff. In this work we choose the cutoff as 0.36 nm, the first minimum position in the $Na^+$-$Cl^-$ RDF. Each bond $\vec{r}$ is associated with a local bond order parameter

$$Q_{lm}(\vec{r}) = Y_{lm}(\theta(\vec{r}), \phi(\vec{r})) \tag{6}$$

where $Y_{lm}(\theta(\vec{r}), \phi(\vec{r}))$ is the spherical harmonic function, $\theta(\vec{r})$ and $\phi(\vec{r})$ are the polar and azimuthal angles of the bond $\vec{r}$ with respect to a chosen fixed reference frame. In our calculations, we only consider the even-$l$ bond order parameters because they are independent of the specific choice of the reference frame. The global bond order parameter is defined as the average over all bonds

$$\overline{Q_{lm}} = \frac{1}{N_b}\sum_{\vec{r}} Q_{lm}(\vec{r}) \tag{7}$$

where $N_b$ is the number of bonds and the summation runs over all bonds. To make the bond order parameter rotationally invariant, it is useful to consider the combinations of

$$Q_l = \left( \frac{4\pi}{2l+1} \sum_{m=-l}^{l} |\overline{Q_{lm}}|^2 \right)^{1/2} \tag{8}$$

and

$$W_l = \sum_{\substack{m_1,m_2,m_3 \\ m_1+m_2+m_3=0}} \begin{pmatrix} l & l & l \\ m_1 & m_2 & m_3 \end{pmatrix} \overline{Q_{lm_1}} \, \overline{Q_{lm_2}} \, \overline{Q_{lm_3}} \tag{9}$$

where the coefficients $\begin{pmatrix} l & l & l \\ m_1 & m_2 & m_3 \end{pmatrix}$ are the Wigner 3$j$ symbol. It is more useful to use the normalized quantity

$$W_{ln} = \frac{W_l}{\left( \sum_m |\overline{Q_{lm}}|^2 \right)^{3/2}} \tag{10}$$

It is generally sufficient to determine a crystal structure by the combination of four bond order parameters: $Q_4, Q_6, W_{4n}$ and $W_{6n}$, whose values for several crystalline structures are listed in Table 2.

**Table 2.** Bond order parameters for different structures including the liquid state, simple cubic (sc), body-centered cubic (bcc), face-centered cubic (fcc), and hexagonal-close-packed (hcp).

| Geometry | $Q_4$ | $Q_6$ | $W_{4n}$ | $W_{6n}$ |
|---|---|---|---|---|
| liquid | 0 | 0 | 0 | 0 |
| sc | 0.76376 | 0.35355 | 0.15932 | 0.01316 |
| bcc | 0.089999 | 0.44053 | 0.15932 | 0.01316 |
| fcc | 0.19094 | 0.57452 | -0.15932 | -0.01316 |
| hcp | 0.09722 | 0.48476 | 0.13410 | -0.01244 |

## 3. Results

In this section, we mainly discuss the influence of $E$ on the nucleation process of ions and the crystallized solution based on our MD simulation results.

### 3.1. Influence of *E* on nucleation

To obtain a panoramic view of the nucleation process with or without *E*, we employed the HOP to describe the nucleation process. A larger HOP value corresponds to a more heterogeneous configuration, indicating that ions aggregate more tightly to form clusters in the nucleation process. The calculated HOPs with time under different *E*s at concentrations $c$ = 5.45 M and 6.24 M are shown in Fig. 1. It can be seen that the applied *E* has a strong influence on the nucleation process. No nucleation or even crystallization happens in the solution when a strong *E* is applied, and the HOPs fluctuate around a small value. The nucleation and crystallization can happen when a moderate *E* is applied, and the HOP initially fluctuates around a mean value and then continuously increase with time. In the situation when crystallization can happen, the onset time $t_s$ for nucleation non-monotonically depends on *E*. For instance, when $c$ = 5.24 M, $t_s$ is around 10 ns at $E$ = 0.1 V/nm, 15 ns at $E$ = 0, and 75 ns at $E$ = 0.15 V/nm.

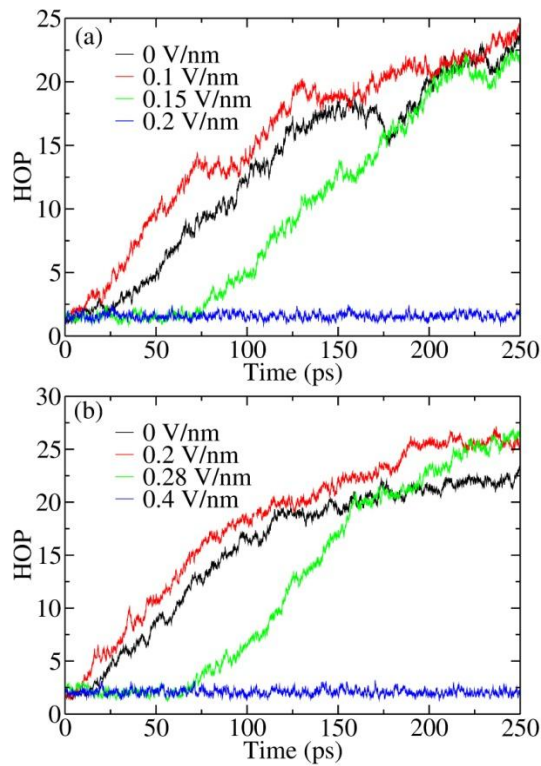

**Fig.1.** HOP evolves with simulation time under different *E*s at $c$ = 5.45M (a) and 6.24M (b), respectively.

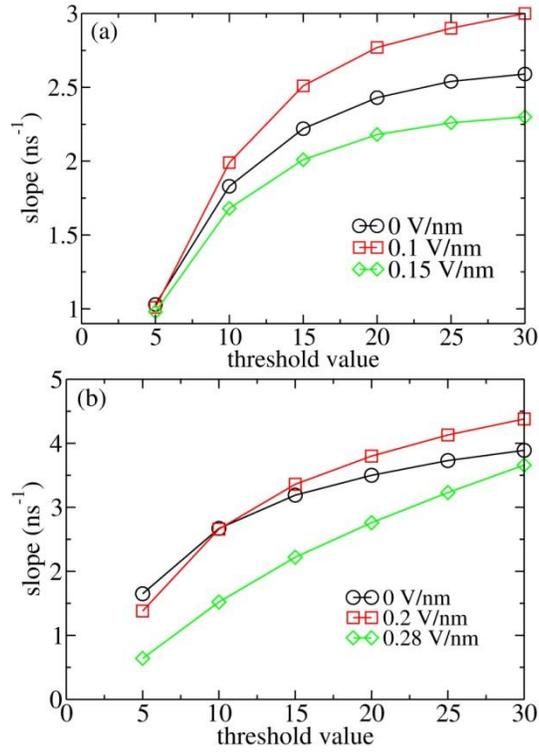

**Fig.2.** Nucleation rates with different threshold values under different $E$s at $c$ = 5.45M (a) and 6.24M (b), respectively.

We employed the threshold method proposed by Yasuoka and Matsumoto to determine the nucleation rates.[39, 40] In this method, the number of ions in a cluster larger than a certain threshold value is calculated versus simulation time, and the nucleation rate defined in the CNT is just the slope of the characteristic linear region of the nucleation growth line divided by the volume. The clusters in our sampled configurations were determined according to Hassan's definition[15] that an ion cluster is an array of ions in which each ion is connected with at least one another ion. Two ions are considered connected if their distance is within a certain cutoff, which was chosen in this work as 0.36 nm, the position of the first minimum of the $Na^+$-$Cl^-$ RDF. The calculated slopes of nucleation at $c$ = 5.45 M and 6.24 M with six different threshold values under three different $E$s are plotted in Fig. 2. We can see that the electric field can either accelerate or decelerate the nucleation process. For instance, at $c$ = 5.45 M, the nucleation rate at $E$ = 0.1 V/nm is greater than $E$ = 0, but the nucleation is decelerated at $E$ = 0.15 V/nm. At the threshold of 20, the nucleation rates are very close to their finally approached values.

Therefore, we have fixed the threshold value to be 20 to calculate the time evolutions of ion numbers in large clusters (larger than 20 ions) and the corresponding curves are plotted in Fig. 3.

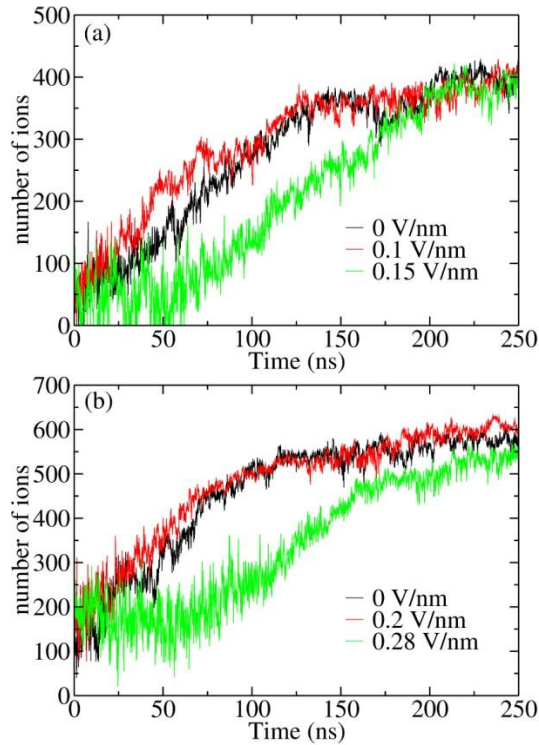

**Fig. 3.** Number of ions in the clusters larger than the threshold value of 20 evolving with simulation time under different $E$s at $c$ = 5.45M (a) and 6.24M (b), respectively.

The general processes of cluster nucleation and growth can be observed in Fig. 3, which have almost the same trend as the HOP shown in Fig. 1. The vibrate fluctuations of the growth lines come from frequent cluster-cluster collisions in a highly saturated solution. Similar to the time evolution of the HOP, the number of ions first fluctuates heavily but shows no apparent increase before it grows almost linearly with time. As time goes on, the nucleation rate decreases because ions gradually deplete from the solution to crystallize. Finally each growth line approaches a plateau with the average number of ions does not change with time, showing that the nucleation process is terminated. The final numbers of ions at the plateau are almost the same, demonstrating that $E$ does not affect the degree of nucleation. From Figs. 3(a) and (b) we can also estimate the onset time of nucleation, $t_s(E)$, which exhibits the same sequence as the HOP shown in Fig. 1: $t_s$ (0.1 V/nm) < $t_s$(0) < $t_s$ (0.15V/nm) at $c$ = 5.45 M, and $t_s$ (0.2V/nm) < $t_s$(0) < $t_s$ (0.28V/nm) at $c$ = 6.24 M.

All the above results indicate that, at $c$ = 5.45 M, the nucleation process at $E$=0.1 V/nm is easier than without $E$ since the nucleation process starts earlier and faster, but at $E$=0.15 V/nm, it starts later and the nucleation rate is smaller than at $E$=0. The same trend is observed for $c$ = 6.24 M. Those results indicate that a suitable $E$ can increase the nucleation rate but a too strong $E$ will decelerate the nucleation process or even no crystallization happens.

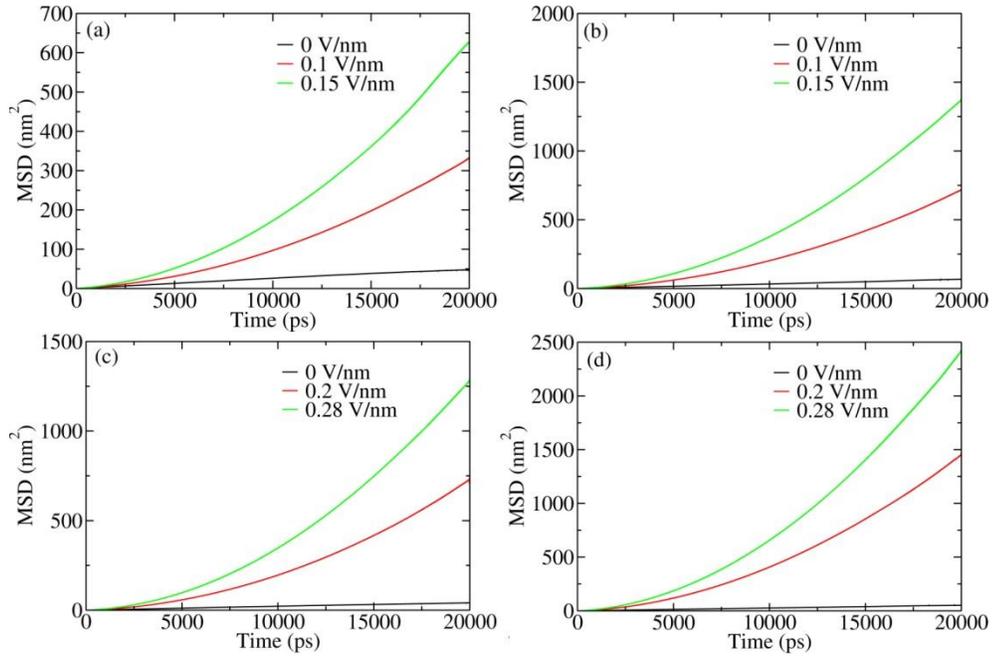

**Fig. 4.** Mean square displacements of (a) Na$^+$ at $c$ = 5.45 M, (b) Cl$^-$ at $c$ = 5.45 M, (c) Na$^+$ at $c$ = 6.24 M, and (d) Cl$^-$ at $c$ = 6.24 M under different $E$s.

To find the microscopic mechanism explaining the above phenomena, we then analyzed ion motions in the NaCl solutions under $E$. The displacement of an ion is composed of thermodynamic motion and drift motion driven by $E$

$$R(t) = R_r(t) + v_d t \qquad (11)$$

where $R(t)$ is the total displacement of an ion, $R_r(t)$ is the displacement due to thermal motion, and $v_d$ is the drift velocity induced by $E$. Taking into account the randomness of the thermal motion $\langle R_r(t) \rangle = 0$ and using the Einstein relation $\langle R_r(t)^2 \rangle = 6Dt$, we obtain the mean square displacement (MSD)

$$\langle R(t)^2 \rangle = v_d^2 t^2 + 6Dt \qquad (12)$$

where $D$ is the diffusion coefficient. The calculated MSDs are shown in Fig. 4 and the fitted ion drift velocities and diffusion coefficients are listed in Tables 3 and 4 for $c$ = 5.45 M and 6.24 M, respectively.

**Table 3.** Diffusion coefficients and drift velocities of Na$^+$ and Cl$^-$ under different $E$s at $c$ = 5.45M.

| $E$ (V/nm) | $D_{Na+}$ ($10^{-3}$ nm$^2$/ps) | $D_{Cl-}$ ($10^{-3}$ nm$^2$/ps) | $v_{Na+}$ ($10^{-3}$ nm/ps) | $v_{Cl-}$ ($10^{-3}$ nm/ps) |
|---|---|---|---|---|
| 0 | 0.420 | 0.556 | 0 | 0 |
| 0.1 | 0.465 | 0.755 | 0.831 | -1.250 |
| 0.15 | 0.499 | 1.050 | 1.191 | -1.771 |

**Table 4.** Diffusion coefficients and drift velocities of Na$^+$ and Cl$^-$ under different $E$s at $c$ = 6.24M.

| $E$ (V/nm) | $D_{Na+}$ ($10^{-3}$ nm$^2$/ps) | $D_{Cl-}$ ($10^{-3}$ nm$^2$/ps) | $v_{Na+}$ ($10^{-3}$ nm/ps) | $v_{Cl-}$ ($10^{-3}$ nm/ps) |
|---|---|---|---|---|
| 0 | 0.354 | 0.437 | 0 | 0 |
| 0.20 | 0.485 | 1.389 | 1.291 | -1.796 |
| 0.28 | 0.826 | 1.667 | 1.726 | -2.357 |

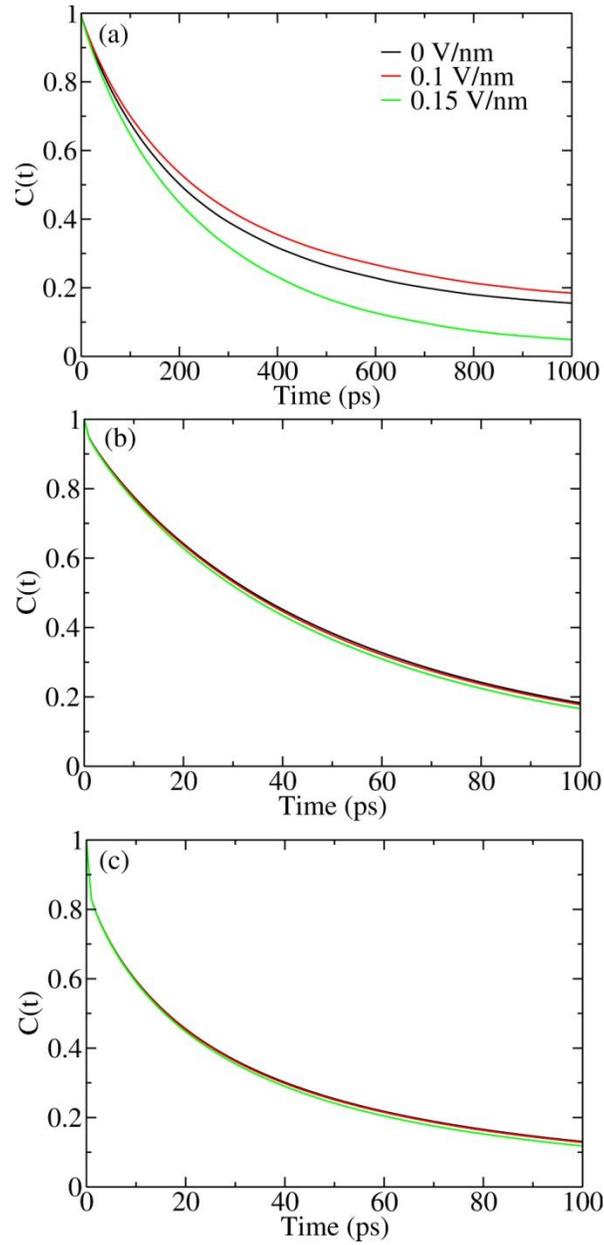

**Fig. 5.** Residence correlation time functions at $c$ = 5.45M under different $E$s: (a) Na$^+$-Cl$^-$, (b) Na$^+$-H$_2$O, and (c) Cl$^-$-H$_2$O.

As can be seen from those data, ion diffusion and drift velocity increase with $E$, whose mechanism has been given in ref 24. Since ions are solvated and surrounded by water molecules, they have to pass through the solvation shell composed of those water molecules to associate and nucleate.[41] The increase in ion diffusion and drift motion can promote ions transport through water shells and thus accelerates the nucleation process because the nucleation rate is governed by diffusion. On the other hand, as shown in Tables 3 and 4, Na$^+$ and Cl$^-$ move in opposite directions, which makes ions difficult to associate. Therefore, the observed non-monotonically influence of $E$ to nucleation rate results from the competition between random self-diffusion and ion drift in opposite directions.

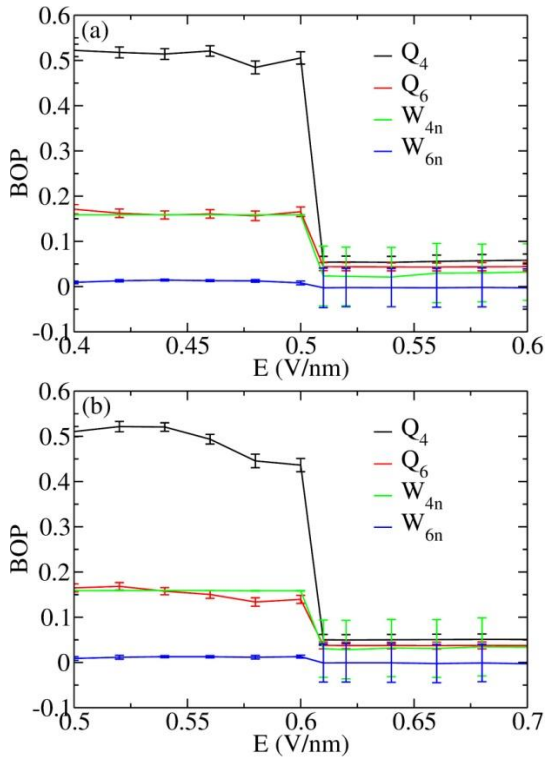

**Fig. 6.** Bond order parameters for ions at $c$ = 5.45 M (a) and 6.24 M (b), respectively.

To further validate the above mechanism, we calculate the residence correlation time functions defined as $C(t) = \langle p(0) p(t) \rangle$, where $p(t)$ is 1 if a given ion is still in the same ion shell at time $t$ and 0 otherwise, and the angular bracket represents the ensemble average over all sampled trajectories. The $C(t)$ for Na$^+$-Cl$^-$, Na$^+$-H$_2$O, and Cl$^-$-H$_2$O at $c$ = 5.24M are plotted in Fig. 5, which characterize the lifetime and stability of a solvation shell from a dynamic point of view. We can see that the Na$^+$-Cl$^-$ stability with respect to $E$ has the tendency 0.1 V/nm > 0 > 0.15 V/nm, and the ion solvation shell stability obeys 0 > 0.1 V/nm > 0.15 V/nm, consistent with the change of the nucleation rate with respect to $E$ and the above mechanism. Those results are different from the prediction by Onsager and Kim's theory[23] and our previous simulations[24] conducted for concentrated but not saturated solutions.

### 3.2. Influence of $E$ on crystallization

In this section we explore the influence of applied $E$ on crystallized NaCl solutions. After crystallization, a certain amount of ions associate to form ordered structures, with Na$^+$ or Cl$^-$ ions forming the face-centered cubic (FCC) crystalline structure, and overall a simple cubic (SC) crystalline structure. Onsager and Kim's theory[23] indicates that the associated ion pairs in saturated solutions tend to dissociate whose degree is almost proportional to $E$. If this theory can also be applied to the crystallization case, the formed crystals have to re-dissolve in water with an $E$ applied. We employed the BOPs to quantify the influence of $E$ on the solutions after crystallization at $c$ = 5.45 M and 6.24 M, respectively. The calculated BOPs for ions are shown in Figs. 6 and 7.

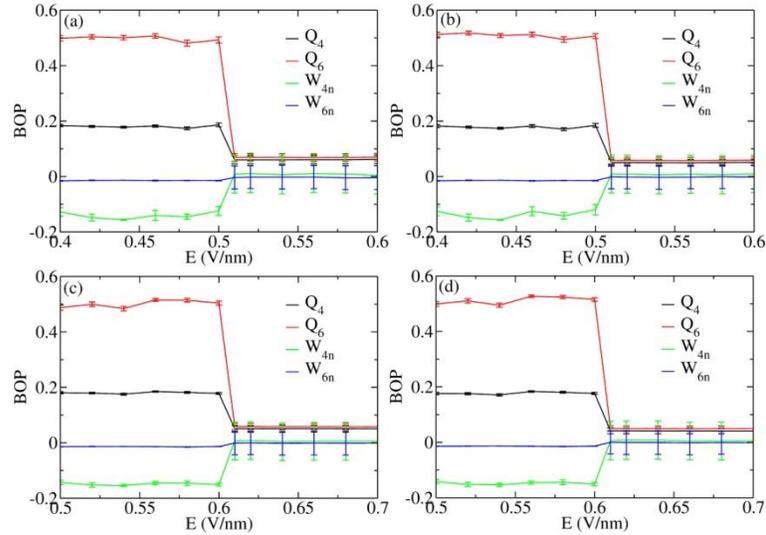

**Fig. 7.** Bond order parameters for different ion species at different concentrations. (a) Sodium at 5.45 M. (b) Chloride at 5.45 M. (c) Sodium at 6.24 M. (d) Chloride at 6.24 M.

The calculated BOPs demonstrate that Na$^+$ or Cl$^-$ ions indeed form FCC and they overall form SC. Because of the limited simulation size and ions on the interface, the calculated BOP values slightly deviate from the ideal values listed in Table 2. With increasing $E$, the BOPs remain almost unchanged up to a certain $E_c$, when they abruptly change to a near-zero value and keep almost unchanged afterwards, indicating a first-order phase transition. Therefore, Onsager and Kim's theory cannot be applied to crystallized solutions, since their theory predicts a continuous rather than an abrupt change of the structure with $E$. The system potential energy as a function of $E$ has also been calculated and is shown in Fig. 8. The potential energy also has an abrupt change at $E_c$, characterizing a first-order phase transition.

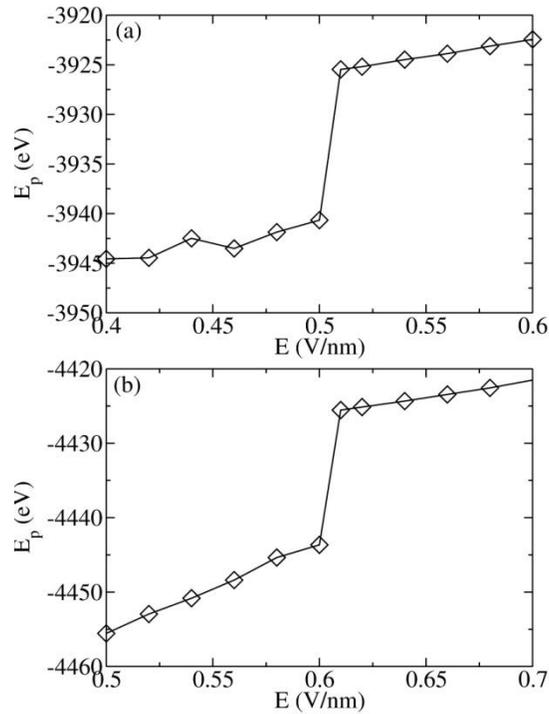

**Fig. 8.** Potential energy as a function of $E$ at $c$ = 5.45 M (a) and 6.24 M (b), respectively.

We further determined from the simulation data that $E_c \approx 0.5$ V/nm for $c$ = 5.45M and $E_c \approx 0.6$ V/nm for $c$ = 6.24 M. This tendency can be explained by investigating cluster sizes in the solutions. As shown in Fig. 9, larger clusters form in the solution at $c$ = 6.24 M than at 5.45M. Since the electrostatic interaction in a larger cluster is stronger than in a smaller cluster, a larger $E$ is required to dissociate a larger cluster. Consequently, $E_c$ is larger for a more concentrated solution, which can be seen from Fig. 10.

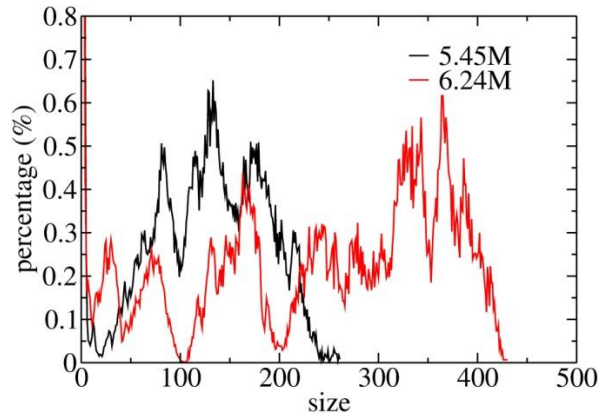

**Fig. 9.** Cluster distributions for the NaCl solutions at $c$ = 5.45M and 6.24M, respectively.

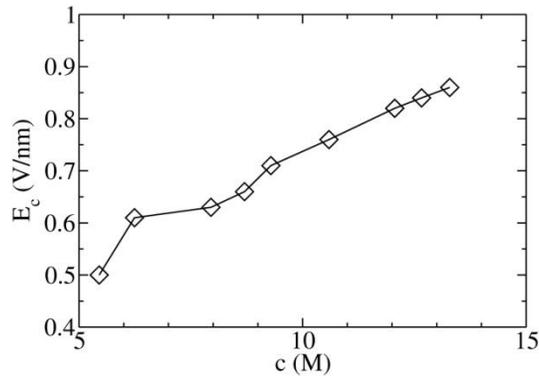

**Fig. 10.** Transition point $E_c$ with respect to ionic solution concentration.

**4. Conclusions and Discussion**

In conclusion, we have carried out a series of non-equilibrium MD simulations for saturated NaCl solutions to investigate the influence of an external static electric field on their nucleation process and crystallization phase transition. We have found that the external field $E$ can either accelerate or decelerate the nucleation process depending on the strength of $E$. For a moderate $E$, the nucleation process starts earlier, progresses faster, and ions are more likely to associate to form stable structures than without $E$, which is different from Wien's observation[22] and our previous simulations for unsaturated solutions.[24] In contrast, a strong $E$ retards the nucleation process or even prevents crystallization. The degree of nucleation does not change with $E$ when crystallization can happen. Those observations can be understood by considering the competition between ion self-diffusion and drift motion. When an $E$ is applied, the increased random self-diffusion and drift motion allow ions to pass through solvation shells and associate more easily, but at the same time the increase of drift motion also makes the oppositely charged ions more difficult to associate. The competition between those two factors result in the non-monotonic change of the nucleation speed with respect to $E$.

The BOPs and the potential energy calculated for the configurations sampled from the steady-state MD simulations exhibit an abrupt change at the transtion point $E_c$, indicating a first-order phase transition. A more concentrated solution has a larger $E_c$, which can be explained by the fact that larger clusters can easily form in a more concentrated solution, so a larger $E$ is required to dissociate them. The observed phenomena and proposed mechanisms are helpful for the wide applications of ionic solutions in various chemical and biological systems.

## References

[1]  Kashchiev D 1982 J. Chem. Phys.  76 5098
[2]  Ford I J 1996 J. Chem. Phys.  105 8324
[3]  Das S, Kear B and Adam C 1985 Morristown  1985
[4]  Toner M, Cravalho E G and Karel M 1990 J. Appl. Phys.  67 1582
[5]  McDonald J E 1962 American Journal of Physics  30 870
[6]  McDonald J E 1963 American Journal of Physics  31 31
[7]  Gan R and Yanting W 2014 EPL (Europhysics Letters)  107 30005
[8]  Robinson R A and Stokes R H, Electrolyte Solutions.Dover Publications, New York, 2002.
[9]  Georgalis Y, Kierzek A M and Saenger W 2000 J. Phys. Chem. B  104 3405
[10]   Bian H, Wen X, Li J, Chen H, Han S, Sun X, Song J, Zhuang W and Zheng J 2011 Proc. Natl. Acad. Sci. U. S. A.  108 4737
[11]  Chialvo A A and Simonson J M 2007 J. Mol. Liq.  134 15
[12]  Fennell C J, Bizjak A, Vlachy V and Dill K A 2009 J. Phys. Chem. B  113 6782

[13]  Sherman D M and Collings M D 2002 Geochem. Trans.  3 102
[14]  Hassan S A 2008 J. Phys. Chem. B  112 10573
[15]  Hassan S A 2008 Phys. Rev. E  77 031501
[16]  Zahn D 2004 Phys. Rev. Lett.  92 040801
[17]  Zahn D 2007 J. Phys. Chem. B  111 5249
[18]  Yang Y and Meng S 2007 J. Chem. Phys.  126 044708
[19]  Gebauer D, Völkel A and Cölfen H 2008 Science  322 1819
[20]  Demichelis R, Raiteri P, Gale J D, Quigley D and Gebauer D 2011 Nat. Commun.  2 590
[21]  Wallace A F, Hedges L O, Fernandez-Martinez A, Raiteri P, Gale J D, Waychunas G A, Whitelam S, Banfield J F and De Yoreo J J 2013 Science  341 885
[22]  Wien M 1928 Ann. d. Physik  85 795
[23]  Onsager L and Kim S K 1957 The Journal of Physical Chemistry  61 198
[24]  Ren G, Shi R and Wang Y 2014 J. Phys. Chem. B  118 4404
[25]  Moroni D, ten Wolde P R and Bolhuis P G 2005 Phys. Rev. Lett.  94 235703
[26]  Lechner W, Dellago C and Bolhuis P G 2011 Phys. Rev. Lett.  106 085701
[27]  Oxtoby D W 1992 J. Phys.: Condens. Matter  4 7627
[28]  Brooks B R, Bruccoleri R E, Olafson B D, Swaminathan S and Karplus M 1983 J. Comput. Chem.  4 187
[29]  Jorgensen W L, Chandrasekhar J, Madura J D, Impey R W and Klein M L 1983 J. Chem. Phys.  79 926
[30]  Berendsen H J C, van der Spoel D and van Drunen R 1995 Comput. Phys. Commun.  91 43
[31]  Van Der Spoel D, Lindahl E, Hess B, Groenhof G, Mark A E and Berendsen H J 2005 J. Comput. Chem.  26 1701
[32]  Nosé S 1984 J. Chem. Phys.  81 511
[33]  Hoover W G 1985 Phys. Rev. A  31 1695
[34]  Essmann U, Perera L, Berkowitz M L, Darden T, Lee H and Pedersen L G 1995 J. Chem. Phys.  103 8577
[35]  Wang Y and Voth G A 2006 J. Phys. Chem. B  110 18601
[36]  Deng L, Wang Y and Ou-yang Z-c 2012 J. Phys. Chem. B  116 10135
[37]  Steinhardt P J, Nelson D R and Ronchetti M 1983 Phys. Rev. B  28 784
[38]  Lechner W and Dellago C 2008 J. Chem. Phys.  129 114707
[39]  Yasuoka K and Matsumoto M 1998 J. Chem. Phys.  109 8451
[40]  Romer F and Kraska T 2007 J. Chem. Phys.  127 234509
[41]  Nielsen A E, Kinetics of precipitation.Pergamon Press Oxford, 1964.